\documentclass[aps,showpacs,pra,twocolumn,superscriptaddress,10pt]{revtex4-2}
\usepackage{graphicx}
\usepackage{dcolumn}
\usepackage{bm}
\usepackage{physics}
\usepackage{amssymb}
\usepackage{textcomp}
\usepackage{amsthm}
\usepackage{mathtools}
\usepackage{empheq}
\usepackage[utf8]{inputenc}
\usepackage[czech,polish,english]{babel}
\usepackage{dsfont}
\usepackage[shortlabels]{enumitem}
\usepackage[T1]{fontenc}
\usepackage{xcolor}
\usepackage{tabularx}
\usepackage{orcidlink}
\usepackage{hyperref}
\hypersetup{
    colorlinks,
    citecolor=blue,
    filecolor=black,
    linkcolor=blue,
    urlcolor=blue
}

\newcommand{\der}{\mathrm{d}}
\DeclareMathOperator{\sinc}{sinc}
\renewcommand{\Re}{\mathrm{Re}}
\renewcommand{\Im}{\mathrm{Im}}

\usepackage[normalem]{ulem}
\newcommand{\stkout}[1]{\ifmmode\text{\sout{\ensuremath{#1}}}\else\sout{#1}\fi}

\begin{document}

\title{Strongly coupled atom--cavity systems under boundary modulation: \\simulating gravitational-wave effects}
\author{Patryk Michalski\orcidlink{0009-0009-0305-7356}}
\email{pmichalski@cft.edu.pl}
\affiliation{Institute of Theoretical Physics, University of Warsaw, Pasteura 5, 02-093 Warsaw, Poland}
\affiliation{Center for Quantum-Enabled Computing, Center for Theoretical Physics, Polish Academy of Sciences, al. Lotnik\'{o}w 32/46, 02-668 Warsaw, Poland}

\author{Jerzy Paczos\orcidlink{0000-0002-0674-9819}}
\email{jerzy.paczos@fysik.su.se}
\affiliation{Department of Physics, Stockholm University, SE-106 91 Stockholm, Sweden}

\author{Navdeep Arya\orcidlink{0000-0003-0730-4835}}
\email{navdeep.arya@fysik.su.se}
\affiliation{Department of Physics, Stockholm University, SE-106 91 Stockholm, Sweden}

\author{Magdalena Zych\orcidlink{0000-0002-8356-7613}}
\email{magdalena.zych@fysik.su.se}
\affiliation{Department of Physics, Stockholm University, SE-106 91 Stockholm, Sweden}

\date{\today}

\begin{abstract}
    One of the proposed platforms in which both quantum and general relativistic effects can become observable is an atom interacting with the electromagnetic field in a gravitational-wave background. The periodic modulation of field modes induced by variations of the spacetime metric modifies the atomic emission spectrum. Notably, the temporal modulation of the mode-frequency induced by a plane gravitational wave can be simulated through modulated boundary conditions, such as moving cavity mirrors. We analyze the impact of this modulation on atom--field interactions in the strong atom--cavity coupling regime, where Rabi oscillations occur. We show analytically that the modulation is resonantly enhanced, leading to measurable imprints in the atomic transition probability. This establishes a realistic and experimentally accessible platform for probing analogue general relativistic effects in quantum optical systems.
\end{abstract}

\maketitle

\section{Introduction}

Among the most fundamental quantum processes are atomic emission and absorption of radiation~\cite{Dirac1927}. The dynamics of these processes depend on the mode structure of the electromagnetic field, which is sensitive to changes in spacetime geometry~\cite{Maybee2019} and boundary conditions~\cite{Gerry2004,Moore1970}. Atom-field systems have been actively explored as probes of noninertial effects~\cite{Ferreri2019,Kozdon2018}, while recent theoretical studies have extended this paradigm to investigate how gravitational waves (GWs) influence light--matter interactions~\cite{Arya2024,Paczos2026}. Observing GW-induced modifications of atomic transition rates or emission spectra would be of foundational interest as a phenomenon at the interface of quantum theory and general relativity. The magnitude of any GW effect on an Earth-based experiment is expected to be very small, and even when the required sensitivity is reached, before one can claim a detection, the physical mechanism should be tested in a controlled setting. A natural way to achieve this in the context of GW effects on atomic emission is to consider an experimentally accessible analogue in which the field modes are modulated not by spacetime curvature, but by time-dependent boundary conditions, which is the focus of this work.

The sensitivity of atomic dynamics to the underlying field structure makes it possible to tailor and control radiative transitions by engineering the field modes accessible to the atom, for example by placing the atom between two mirrors of an optical cavity. In the weak-coupling regime, where interaction strength is smaller than the decay rates of the atomic polarization and the cavity mode to which the atom couples, the resulting enhancement of the atom's spontaneous emission rate is known as the Purcell effect~\cite{Purcell1946}. In the opposite regime, when the coupling is larger than the decay rates, an emitted photon can be reabsorbed by the atom before escaping the cavity, giving rise to Rabi oscillations~\cite{Jaynes1963}. These regimes are central to both theoretical and experimental investigations of coherent light--matter interactions, providing a platform for testing fundamental quantum effects~\cite{Haroche2006} and for implementing protocols in quantum information science~\cite{Reiserer2015,Gonzalez-Tudela2024}.

In this work, we analyze the interaction of a two-level atom with a single cavity mode whose resonance frequency is modulated by motion of one of the cavity mirrors. Such a setup provides an analogue of the GW influence on atom--field coupling for a single field mode in free space. Mirror oscillations induce a weak periodic modulation of the field mode, which is mathematically equivalent to that produced by a plane GW, and the resulting correction to atomic emission rates also parallels recent predictions of GW imprints in spontaneous emission~\cite{Paczos2026}. At the same time, the setup considered here extends the line of research on time-dependent cavity modulation, which is directly relevant to both literal cavity-optomechanical implementations~\cite{Chang2009,Aspelmeyer2014,Holz2015,Wang2026} and circuit quantum electrodynamics (circuit QED) analogs~\cite{Wallraff2004,Zhukov2017,Blais2021}.

The assumptions of our model can be regarded as conservative: the mirror trajectory is treated classically and the modulation is adiabatic, so that the field follows the instantaneous cavity frequency without intermode scattering or photon-pair production~\cite{Law1994}. Even within this restrained regime, we find nontrivial and experimentally testable signatures in internal atomic transitions---in particular, a resonantly enhanced correction to the excited-state population dynamics. We combine perturbative and multiple-scales analyses to identify the optimal sensing window and parameter regimes. To assess the extractable information, we consider the estimation of the mirror's modulation amplitude and analyze the classical Fisher information associated with atomic population measurements. The parameter requirements single out circuit QED as the most feasible platform for implementing cavity analogue of GW effects on atomic emission.

\section{Setup}\label{sec:setup}

\begin{figure}[h]
  \centering
  \includegraphics[width=\linewidth]{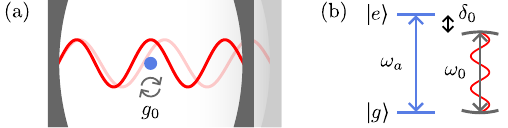}
  \caption{(a) Schematic depiction of the considered cavity with a two-level atom inside. The right mirror can oscillate harmonically. (b) Atomic and cavity mode energy level scheme.}
  \label{fig:scheme}
\end{figure}

We consider a two-level atom interacting with a single quantized mode of the electromagnetic field in a cavity, as depicted in Fig.~\ref{fig:scheme}. The cavity consists of two perfectly reflecting mirrors, one of which is movable and undergoes harmonic oscillations with frequency $\omega$. At time $t$, the mirrors are separated by a distance $L(t) = L_0[1 + \mathcal{A} \cos(\omega t + \phi)]$, where $L_0$ denotes the equilibrium cavity length, $\mathcal{A}$ is the small modulation amplitude and $\phi$ is the initial phase of the oscillations. Due to varying boundary conditions imposed by the movement of the mirror, the instantaneous cavity resonance frequency changes in time according to $\omega_c(t) = \omega_0 L_0/L(t)$, with $\omega_0$ denoting the equilibrium resonance frequency.

For $\omega L_0/\pi c\ll 1$, where $c$ is the speed of light, retardation effects due to finite light-propagation time across the cavity are negligible. In this regime, delayed field response, intermode scattering, and modulation-induced photon-pair creation are suppressed, so the frequency of the electromagnetic field adiabatically follows the mirror motion and is well approximated by the instantaneous expression $\omega_c(t)$~\cite{Law1994}. This behavior is analogous to the modulation of the field mode frequency induced by a plane GW in free space.

The atom is treated as a point-like two-level system with ground state $\ket{g}$, excited state $\ket{e}$, and energy gap $\hbar\omega_a$. We assume that it is initially placed at an antinode of the electromagnetic field, where the atom--field coupling is maximal. Mirror motion shifts the antinode position by an amount proportional to the modulation amplitude $\mathcal{A}$. The atom is assumed to remain fixed rather than following this shift, a situation realized in setups where the atomic trap is independent of the cavity~\cite{Liu2023,Yan2023}. The local field amplitude sampled by the atom decreases only quadratically with the displacement from the antinode. This correction is therefore of second order in $\mathcal{A}$ and can be neglected.

Let us introduce the atomic inversion operator $\hat{\sigma}_z \equiv \ketbra{e}{e} - \ketbra{g}{g}$, atomic raising and lowering operators, $\hat{\sigma}_+ \equiv \ketbra{e}{g}$ and $\hat{\sigma}_- \equiv \ketbra{g}{e}$, as well as annihilation and creation operators of the cavity mode, $\hat{a}$ and $\hat{a}^\dag$. The total Hamiltonian for the system is given by:
\begin{equation}
    \hat{H}_\mathrm{s}(t) = \hat{H}_\mathrm{atom} + \hat{H}_\mathrm{field}(t) + \hat{H}_\mathrm{int}(t),
\end{equation}
with three terms describing the two-level atom, the cavity field, and the atom--field interaction, respectively, all specified in the Jaynes-Cummings model~\cite{Gerry2004,Grynberg2010}:
\begin{align}
    &\hat{H}_\mathrm{atom} = \hbar \omega_a \frac{\hat{\sigma}_z}{2}, \\
    &\hat{H}_\mathrm{field}(t) = \hbar \omega_c(t)\, \hat{a}^\dagger \hat{a}, \\
    &\hat{H}_\mathrm{int}(t) = - \hat{d} \hat{E}(t).
\end{align}
Here, $\hat{d} = d(\hat{\sigma}_+ + \hat{\sigma}_-)$ is the atomic dipole operator, and $\hat{E}(t)$ is the electric field operator at the atomic position, given by~\cite{Lambropoulos2006}:
\begin{equation}
    \hat{E}(t) = \sqrt{\frac{\hbar \omega_c(t)}{\varepsilon_0 A L(t)}} \left(\hat{a} + \hat{a}^\dag\right) = \frac{E_0 L_0}{L(t)} \left(\hat{a} + \hat{a}^\dag\right),
\end{equation}
where $A$ is the transverse mode area, $\varepsilon_0$ is the vacuum permittivity, and $E_0$ is the equilibrium amplitude of the field. Employing
the rotating wave approximation, which holds for small atom--field detuning $\delta(t) \equiv \omega_a - \omega_c(t)$, the interaction
term simplifies to:
\begin{equation}\label{eq:H_int}
  \hat{H}_\mathrm{int}(t) = \hbar g_c(t) \left(\hat{\sigma}_+\hat{a} + \hat{\sigma}_-\hat{a}^\dag\right),
\end{equation}
with the coupling coefficient $g_c(t) = - d E_0 L_0/[\hbar L(t)]$ governing the strength of the atom-light interaction. Assuming that the equilibrium coupling constant $g_0 = -dE_0/\hbar$ satisfies $g_0 \ll \omega_0$, the amplitude of the coupling coefficient modulation, $\mathcal{A}g_0$, is negligibly small as compared to the amplitude of the cavity frequency modulation, $\mathcal{A}\omega_0$. The time dependence of $g_c(t)$ therefore gives only a subleading correction to the dynamics, and we may set $g_c(t) = g_0$ to leading order.

Notably, the simplified Hamiltonian commutes with the total-excitation-number operator $\hat{N} = \ketbra{e}{e} + \hat{a}^\dag \hat{a}$, which implies conservation of the total number of excitations. Therefore, the Hilbert space decomposes into invariant two-dimensional subspaces labeled by the excitation number $n \geq 1$. Each subspace is spanned by $\{\ket{e,n-1},\ket{g,n}\}$, where $\ket{n}$ denotes a photon-number state. In the following sections, we restrict the analysis to a fixed $n$-excitation subspace. More specifically, we assume that the system is initialized in the state $\ket{g,n}$, i.e., the atom is initially in the ground state and the field is in the $n$-th Fock state. Owing to the block structure of the Hilbert space, the extension to other initial field states, such as coherent states, is straightforward.

For convenience, we move to a frame whose phase accumulates at the instantaneous cavity frequency, i.e., with rotation angle $\int_0^t \omega_c(t')\der t'$, for both the atom and the field. In this rotating frame, the Hamiltonian takes the form:
\begin{align}\label{eq:Hamiltonian}
  \hat{H}(t) = \hbar \delta(t) \frac{\hat{\sigma}_z}{2} + \hbar g_0 \left( \hat{\sigma}_+\hat{a} + \hat{\sigma}_-\hat{a}^\dag\right).
\end{align}
This transformation removes the optical phase factors that are common within each $n$-excitation subspace, leaving the dynamics governed by the instantaneous detuning $\delta(t)$ and the coupling strength $g_0$. 

Periodic modulation of the field frequency, incorporated into the Hamiltonian~\eqref{eq:Hamiltonian} through $\delta(t)$, reproduces the effect of a plane GW on a single field mode in free space, where the GW similarly modulates the mode frequency~\cite{Paczos2026}. This modulation causes the evolution of the combined atom--field state to differ from the static equilibrium case. Consequently, the considered configuration is expected to reproduce the essential features of atom--field dynamics in a plane GW background when restricted to a single field mode, which we discuss later in this article.

\section{Perturbative solution}\label{sec:perturbative_solution}

We aim to find a perturbative solution for the evolution of the system described by the Hamiltonian \eqref{eq:Hamiltonian} up to first order in $\mathcal{A}$. Expanding $\delta(t) = \delta_0 + \mathcal{A}\omega_0\cos(\omega t + \phi)$, the Hamiltonian can be split into the time-independent part $\hat{H}_0$ that corresponds to the standard Jaynes--Cummings model and the time-dependent perturbation Hamiltonian $\hat{V}(t)$, given by:
\begin{equation}
  \hat{V}(t) = \mathcal{A} \hbar \omega_0 \frac{\hat{\sigma}_z}{2} \cos(\omega t + \phi).
\end{equation}
The free evolution under $\hat{H}_0$ is exactly solvable, leading to well-known Rabi oscillations~\cite{Gerry2004}. To find the effect of the small modulation, we perform the perturbative expansion. Let us consider the system initially prepared in the state $\ket{\psi_0} = \ket{g,n}$. The state of the system at time $t$ up to first order in $\mathcal{A}$ is:
\begin{equation}
    \ket{\psi(t)} = \hat{U}_0(t) \ket{\psi_0} - \frac{i}{\hbar}\hat{U}_0(t) \int_0^t \der t'\, \hat{V}_\mathrm{I}(t') \ket{\psi_0},
\end{equation}
where $\hat{U}_0(t) = e^{-\frac{i}{\hbar}\hat{H}_0 t}$ is the free evolution operator and $\hat{V}_\mathrm{I}(t) = U_0^\dagger(t)V(t)U_0(t)$ is the perturbation Hamiltonian transformed to the interaction picture. Given this, we calculate the probability $p_e(t) = \abs{\braket{e,n-1}{\psi(t)}}^2$ of finding the system in the state $\ket{e,n-1}$ after a time $t$.

\begin{figure}
    \centering
    \includegraphics[width=\linewidth]{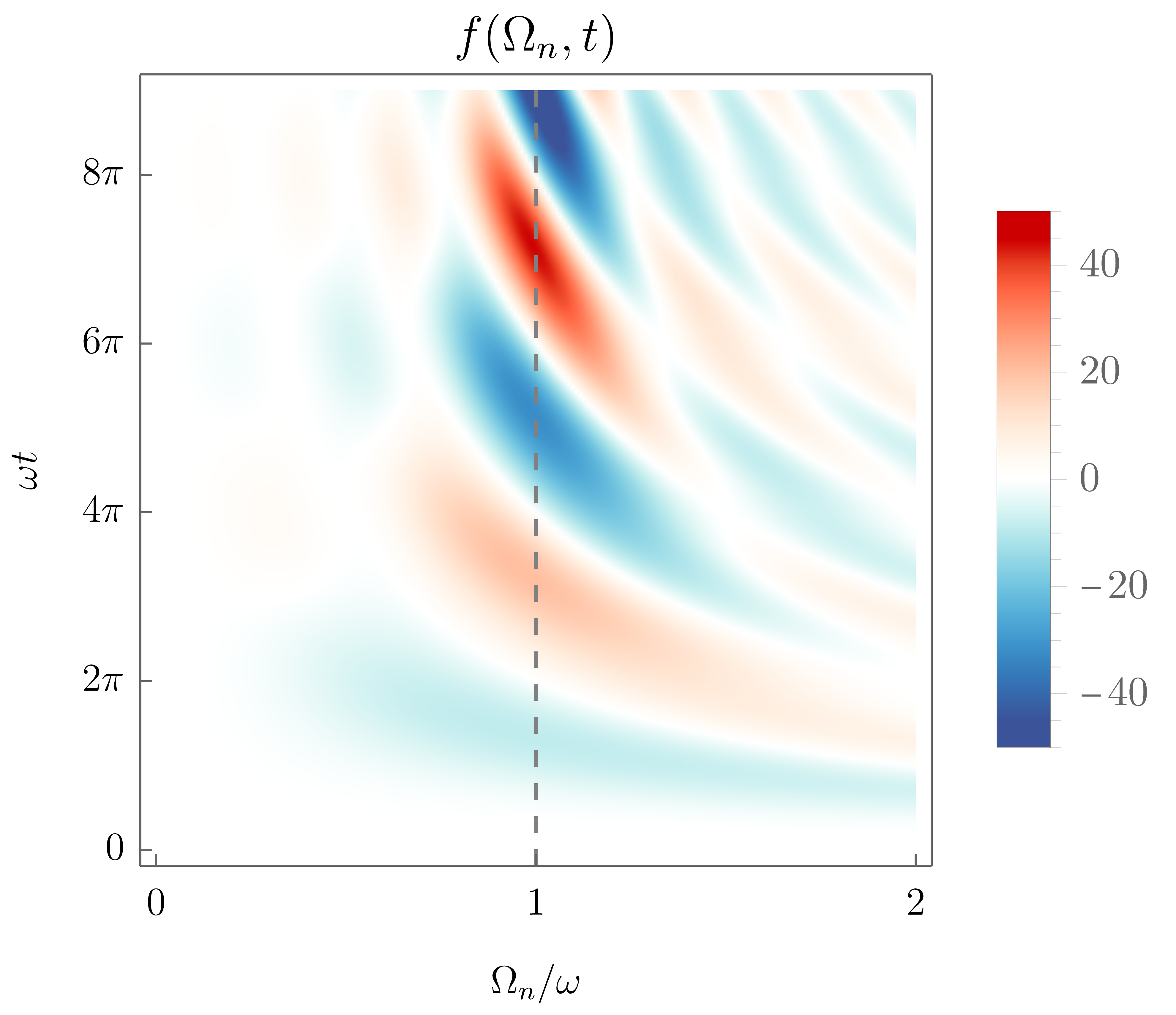}
    \caption{Function $f(\Omega_n,t)$ governing frequency dependence of the probability correction $\delta p_e(t)$. Resonant enhancement for $\Omega_n \approx \omega$ is visible.}
    \label{fig:f}
\end{figure}

In the perturbative expansion, the probability naturally splits, $p_e(t) = \tilde{p}_e(t) + \delta p_e(t)$, into an equilibrium contribution $\tilde{p}_e(t)$ and a correction $\delta p_e(t)$ due to cavity frequency modulation. Denoting the generalized equilibrium Rabi frequency by $\Omega_n \equiv \sqrt{\delta_0^2 + 4g_n^2}$ with $g_n = g_0 \sqrt{n}$ and introducing the ratio $r \equiv \delta_0/\Omega_n$, we can write the respective contributions, up to first order in $\mathcal{A}$, as follows (see Appendix~\ref{Appendix: Perturbative solution}):
\begin{align}
    &\tilde{p}_e(t) = \left(1-r^2\right) \sin^2(\Omega_n t/2),\label{eq:p_e(t)}\\
    &\delta p_e(t) = \mathcal{A} \frac{\omega_0}{\omega} \cos(\omega t/2 + \phi) f(\Omega_n,t) g(r)\label{eq:delta p_e(t)},
\end{align}
where the functions $f(\Omega_n,t)$ and $g(r)$ are given by:
\begin{align}
    &f(\Omega_n,t) \equiv  \Omega_n^2 t^2 \sinc(\Omega_n t/2) \{\sinc[(\Omega_n + \omega)t/2] \nonumber\\
    &\hspace{1.6cm}-\sinc[(\Omega_n - \omega)t/2]\},\\
    &g(r) \equiv r (1-r^2)/4.
\end{align}
The form of $g(r)$ shows that the correction $\delta p_e(t)$ is maximal for $r = \pm 1/\sqrt{3}$, which corresponds to $\delta_0 = \pm \sqrt{2} g_n$. The function $f(\Omega_n,t)$ determines the frequency dependence of the correction. It is plotted in Fig.~\ref{fig:f}, which shows that the effect of mirror vibrations is the strongest near the resonance $\Omega_n \approx \omega$.

The amplitude of the correction $\delta p_e(t)$ is controlled by the factor $\mathcal{A} \omega_0/\omega$. For the perturbative approximation to remain valid, we require this factor to satisfy $\mathcal{A} \omega_0/\omega \ll 1$. At the same time, the ratio $\omega_0/\omega$ can be very large, which amplifies the effect of a small modulation amplitude. Specifically, for mechanical vibrations in the kHz range and $\omega_0/2\pi$ in the microwave range ($\sim 10^{10}$~Hz), one has $\omega_0/\omega \sim 10^{7}$. Note that the same amplification mechanism underlies both interferometric detectors of gravitational waves~\cite{Maggiore2007}, and proposed detection schemes based on gravitational wave imprints on spontaneous emission~\cite{Paczos2026}.

The results in Eqs.~\eqref{eq:p_e(t)}--\eqref{eq:delta p_e(t)} closely resemble the expressions derived in Ref.~\cite{Paczos2026} for spontaneous emission on a GW background; see, in particular, Eq.~(8) in Ref.~\cite{Paczos2026}. For a single field mode, the effect of a plane GW on atom--field dynamics in free space is analogous to that of mirror oscillations in a cavity: in both cases, the quantum field coupled to the atom undergoes periodic modulation. In the cavity case, however, rather than focusing on the directionality of the emission or the sidebands appearing in the emission spectrum, the analogy concerns the modulation-induced correction to the single-mode photon population. In the cavity, this directly corresponds to the correction of the atomic excited-state population. The two descriptions coincide in the dispersive regime $|g_n/\delta_0| \ll 1$ (equivalently, $|r| \to 1$), where one can expand in powers of $g_n$. To lowest order, the dynamics reduces to a single transition rather than coherent atom--field oscillations. Away from this regime, the atom--field coupling modifies the level splitting and replaces the spontaneous-emission detuning $\delta_0$ with the generalized Rabi frequency $\Omega_n$.

In our setup, the effect can be detected by repeated state-selective measurements that search for an excess or deficit in the excited-state population. To maximize sensitivity, one should operate near $\Omega_n\approx\omega$ and choose interrogation times long enough for the correction to accumulate. However, in the resonant limit $\Omega_n \to \omega$, secular terms linear in time appear in the perturbative correction:
\begin{align}\label{eq:perturbative_correction}
  \delta p_e(t) ={} & \mathcal{A} \frac{\omega_0}{\omega} g(r) \left[\sin(\omega t + \phi) - \sin(\phi)\right]\nonumber\\
  &\times \left[\sin(\omega t)-\omega t\right].
\end{align}
This causes the perturbative expansion to break down at times $t \sim (\mathcal{A} \omega_0)^{-1}$. Therefore, to remain within the perturbative regime, the interrogation time should satisfy $t\lesssim(\mathcal{A}\omega_0)^{-1}$.

\section{Amplitude estimation}\label{sec:amplitude_estimation}

To quantify how much information about mirror vibrations can be extracted, we consider estimating the relative displacement amplitude $\mathcal{A}$ from atomic population measurements. The minimal achievable uncertainty $\delta\mathcal{A}$ of such an estimation is related to the classical Fisher information $\mathcal{I}_\mathrm{C}(t)$ associated with the measurement and the number $M$ of independent repetitions of the measurement through $\delta\mathcal{A} \geq 1/\sqrt{M \mathcal{I}_\mathrm{C}(t)}$~\cite{Cramer1946, Rao1945}. Treating $p_i$ as a discrete probability distribution for a two-outcome measurement (with $i\in\{e,g\}$) conditioned on the value of $\mathcal{A}$, we can calculate the classical Fisher information using the standard formula $\mathcal{I}_\mathrm{C}(t) = \sum_i (\partial_\mathcal{A} p_i)^2/p_i$~\cite{Fisher1922}. To leading (zeroth) order in $\mathcal{A}$, this gives:
\begin{align}
    \mathcal{I}_\mathrm{C}(t) = \frac{[\delta p_e(t)/\mathcal{A}]^2}{\tilde{p}_e(t)[1-\tilde{p}_e(t)]},
\end{align}
where $\tilde{p}_e(t)$ is the unperturbed Jaynes--Cummings contribution and $\delta p_e(t)$ is the first-order correction derived in Sec.~\ref{sec:perturbative_solution}.
The denominator reflects the usual binary-outcome variance, while the numerator quantifies the sensitivity of the signal to the perturbation.

Substituting the perturbative expressions from Sec.~\ref{sec:perturbative_solution} in the resonant limit $\Omega_n \to \omega$ gives:
\begin{align}
    \mathcal{I}_\mathrm{C}(t) ={} & \frac{1}{4} \left(\frac{\omega_0}{\omega}\right)^2 h(r,\phi,t)\,[\sin(\omega t) - \omega t]^2,
\end{align}
where the modulating function $h(r,\phi,t)$ is:
\begin{equation}
    h(r,\phi,t) \equiv \frac{r^2\left(1-r^2\right)\cos^2(\omega t/2 + \phi)}{1-\left(1-r^2\right)\sin^2(\omega t/2)}.
\end{equation}
The remaining factor defines a monotonic envelope and makes the enhancement factor $(\omega_0/\omega)^2$ explicit. As shown in Appendix~\ref{Appendix: Fisher information}, the function $h(r,\phi,t)$ attains local maxima at $t_m = 2\{\pi m - \arctan[\tan(\phi)/r^2]\}/\omega$, with $m \in \mathbb{N}$. Although it is not possible to obtain analytically the exact interrogation times that maximize the classical Fisher information $\mathcal{I}_\mathrm{C}(t)$, the times $t_m$ provide an accurate approximation.

\begin{figure}
    \centering
    \includegraphics[width=\linewidth]{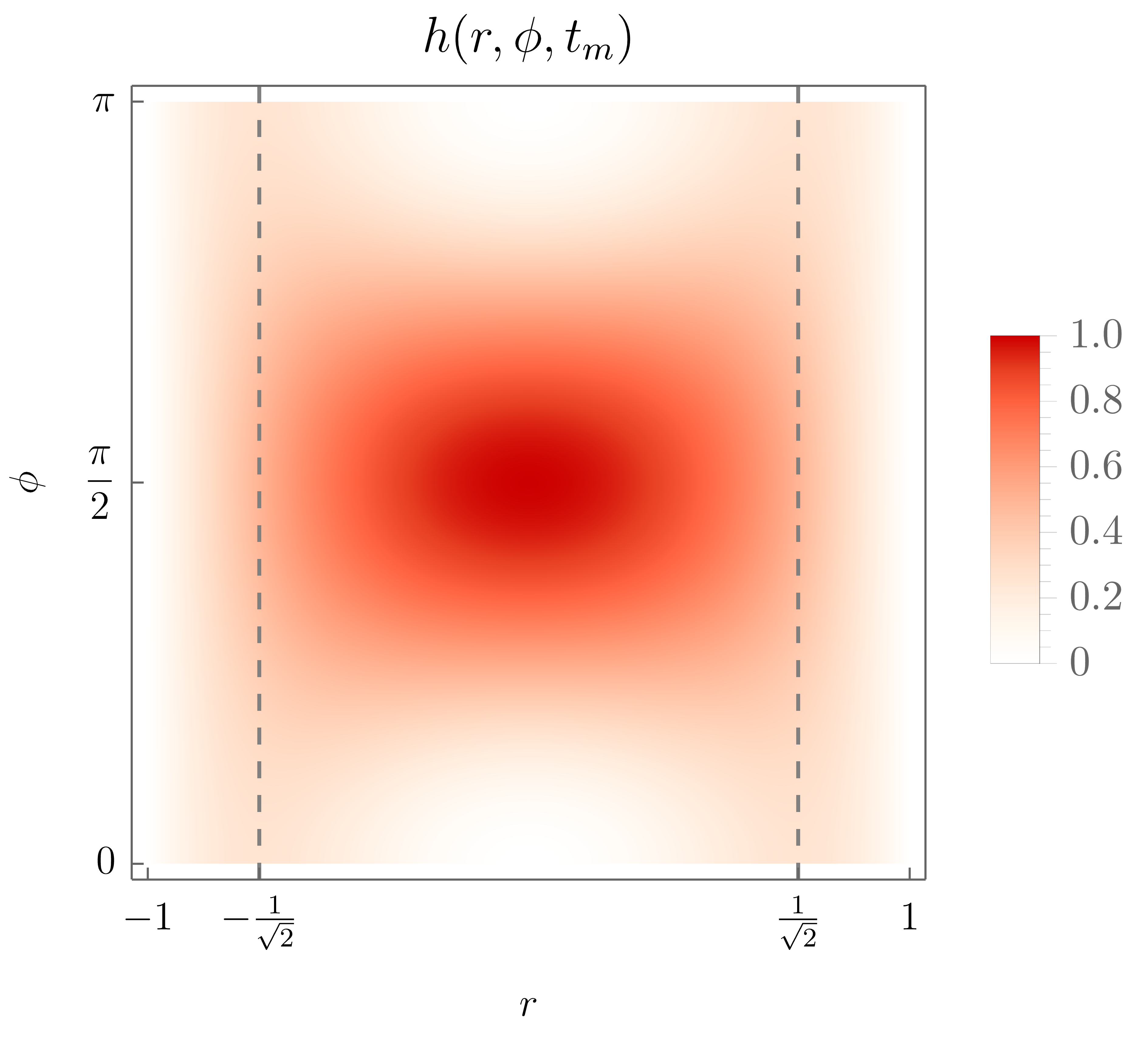}
    \caption{Function $h(r,\phi,t_m)$ determining the magnitude of the classical Fisher information $\mathcal{I}_\mathrm{C}(t)$ at an optimal interrogation time $t_m$. The dashed vertical lines indicate the values $r = \pm 1/\sqrt{2}$ around which the sensitivity is nonvanishing regardless of the initial phase $\phi$.}
    \label{fig:h}
\end{figure}

At any chosen time $t_m$, the function $h(r,\phi,t)$ takes the form:
\begin{equation}
    h(r,\phi,t_m) = (1-r^2)[r^2\cos^2(\phi)+\sin^2(\phi)].
\end{equation}
For $\phi = 0$ or $\phi = \pi$, this expression vanishes at $r \to 0$ and is maximal at $r = \pm 1/\sqrt{2}$. These two values of the ratio $r$ guarantee nonvanishing sensitivity for any value of the initial phase $\phi$ (see Fig.~\ref{fig:h}). They correspond to $\delta_0 = \pm 2g_n$. Equivalently, the detuning $\delta_0$ and the rescaled coupling $g_n$ should contribute comparably to the generalized Rabi frequency. By contrast, when the initial phase can be tuned, the optimal choice that maximizes the Fisher information is $\phi = \pi/2$ or $\phi = 3\pi/2$ together with $r \to 0$, corresponding to vanishing detuning. In this limit, however, the first-order correction to the excited state population, $\delta p_e(t)$, vanishes. It should therefore be regarded as an idealized optimum, because higher-order corrections and experimental noise determine the achievable sensitivity.

Moreover, notice that the Fisher-optimal values of the ratio $r$ differ from the values $r=\pm1/\sqrt{3}$ that maximize the probability correction itself. This is because the Fisher information weights the squared signal by the inverse binary-outcome variance, $\tilde p_e(t)[1-\tilde p_e(t)]$, and therefore favors the choice of parameters where a given correction is more statistically distinguishable rather than merely larger in absolute magnitude.

\section{Multiple-scales analysis}\label{sec:multiple_scales}

For cavities with long coherence times, the effect of field modulation can accumulate beyond the temporal window in which first-order perturbation theory remains reliable. In this regime, a description that captures coupled fast and slow dynamics on long timescales is needed. Near resonance, $\Omega_n \approx \omega$, the multiple-scales method~\cite{Bender2010} provides such an approximation valid beyond $t \sim (\mathcal{A} \omega_0)^{-1}$ by resumming long-timescale linear growth into slowly varying amplitudes.

Let us again consider the system initially prepared in the state $\ket{\psi_0} = \ket{g,n}$. After time $t$, the state remains in the $n$-excitation subspace and can be decomposed as:
\begin{equation}
  \ket{\psi(t)} = c_e(t) \ket{e,n-1} + c_g(t) \ket{g,n}.
\end{equation}
The evolution is governed by the Schr\"odinger equation:
\begin{equation}
  i\hbar\partial_t\ket{\psi(t)} = \hat{H}(t)\ket{\psi(t)},
\end{equation}
which yields the following set of differential equations for the amplitudes $c_e(t)$ and $c_g(t)$:
\begin{subequations}
\begin{empheq}[left=\empheqlbrace]{align}
  i \dot{c}_e(t) &= \tfrac{1}{2} \delta(t) c_e(t) + g_n c_g(t),\\
  i \dot{c}_g(t) &= -\tfrac{1}{2} \delta(t) c_g(t) + g_n c_e(t).
\end{empheq}
\end{subequations}
These equations can be decoupled into second-order ordinary differential equations and solved using the multiple-scales approach. 

We aim to derive the probability of finding the system in the state $\ket{e,n-1}$ after time $t$, which is given by $p_e(t) = \abs{c_e(t)}^2$. To this end, we introduce the parameter $\alpha \equiv \mathcal{A}\omega_0/\Omega_n$, which defines the slow timescale $t_1 = \alpha t$. Since $\Omega_n \approx \omega$, we have $\alpha \ll 1$, consistently with the assumptions from Sec.~\ref{sec:perturbative_solution}. The solution $c_e(t)$ is then assumed to admit a perturbative expansion that depends on both $t$ and $t_1$. Although there is no natural splitting of the obtained probability into free and perturbed parts, one can nevertheless compute the correction $\delta p_e(t)$ by subtracting the equilibrium contribution $\tilde{p}_e(t)$ from the full probability $p_e(t)$. 

The general expression for $c_e(t)$ corresponding to the initial conditions $c_e(0)=0$ and $c_g(0)=1$, obtained via the multiple-scales method, is presented in Appendix~\ref{Appendix: Multiple-scales analysis}. Here, we focus on the resonant case $\Omega_n \to \omega$, for which the probability correction takes the form:
\begin{equation}\label{eq:MS_correction}
    \delta p_e(t) = \mathcal{B}(r,\phi,t) + \mathcal{E}(r,\phi,t) \sin(\omega t + \phi),
\end{equation}
where the baseline $\mathcal{B}(r,\phi,t)$ and the envelope $\mathcal{E}(r,\phi,t)$ oscillate at the slow-timescale frequency $\mu \equiv \alpha g_n$ and are given explicitly by:
\begin{align}
    &\mathcal{B}(r,\phi,t) \equiv r\sqrt{1-r^2} \sin(\mu t/2) \cos(\mu t/2) \sin(\phi) \nonumber\\
    &\hspace{1.7cm}+r^2 \sin^2(\mu t/2),\\
    &\mathcal{E}(r,\phi,t) \equiv (1-r^2) \sin^2(\mu t/2) \sin(\phi) \nonumber\\
    &\hspace{1.7cm}- r\sqrt{1-r^2} \sin(\mu t/2) \cos(\mu t/2).
\end{align}
This solution implicitly incorporates corrections of higher order in $\mathcal{A}$ into the oscillatory terms. For a fixed ratio $r$ and initial modulation phase $\phi$, the quantities $\mathcal{B}(t) \pm |\mathcal{E}(t)|$ specify, respectively, the upper and lower envelopes governing the fast dynamics of the correction.

\begin{figure}
    \centering
    \includegraphics[width=\linewidth]{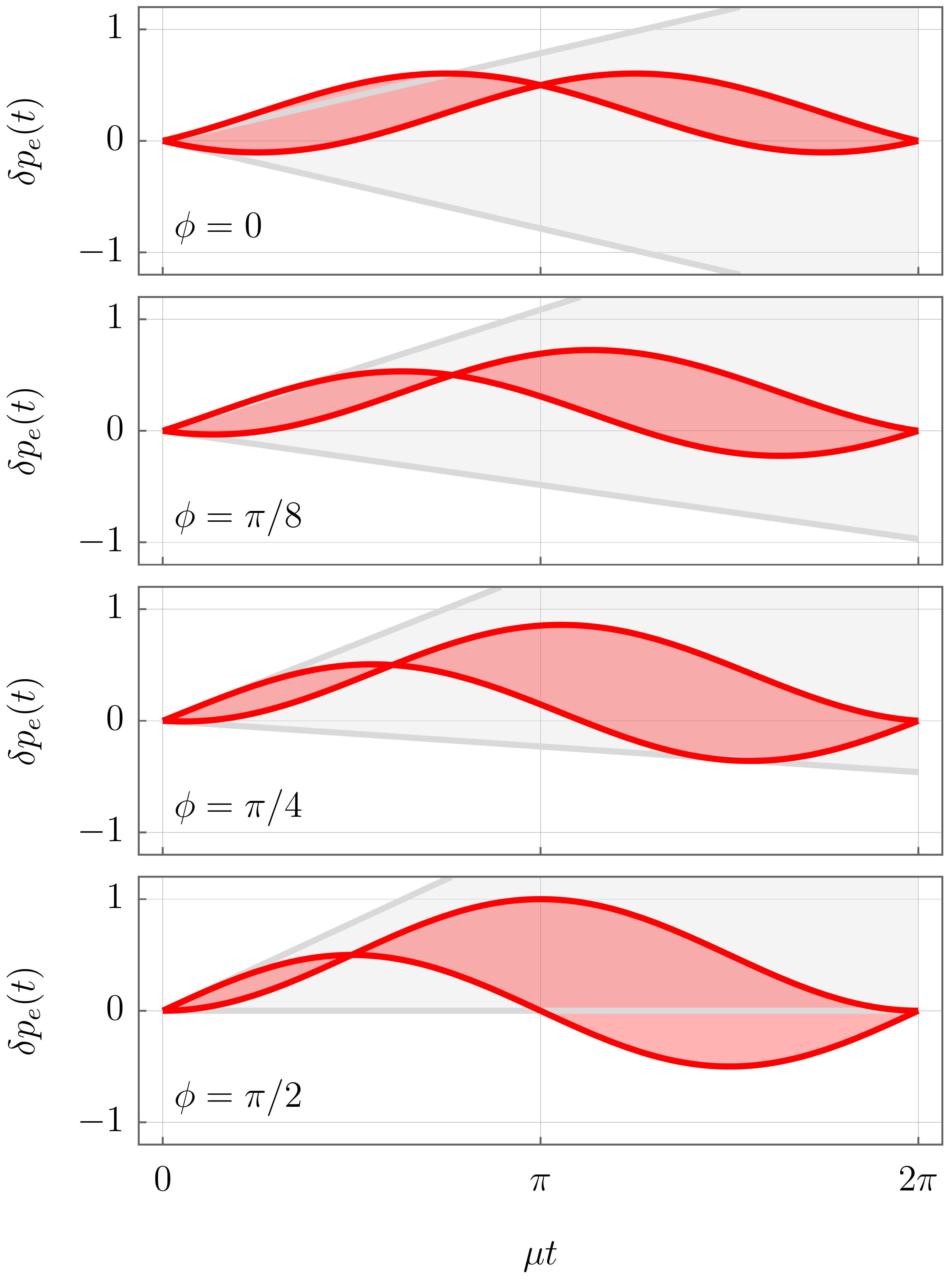}
    \caption{Slow-timescale envelopes of the probability correction $\delta p_e(t)$ in the resonant regime $\Omega_n \to \omega$ for the multiple-scales (in red) and perturbative (in grey) solutions, shown for $r=\pm 1/\sqrt{2}$ and various choices of the initial modulation phase $\phi$. }
    \label{fig:dp}
\end{figure}

In Fig.~\ref{fig:dp}, we plot the slow-timescale envelopes of the probability correction for both the multiple-scales and perturbative solutions, using the ratio $r = \pm 1/\sqrt{2}$ and three distinct choices of the initial phase $\phi$. The breakdown of the perturbative approximation around $t \sim (\mathcal{A} \omega_0)^{-1}$ is clearly visible. In contrast, the multiple-scales solution more accurately tracks the system’s evolution over longer times. Note, however, that to first order in $\alpha$, the expression~\eqref{eq:MS_correction} does not coincide with the perturbative result~\eqref{eq:perturbative_correction}. The reason is that the solvability condition employed in the multiple-scales method omits the effect of higher harmonics in $\omega$, which in any case do not exhibit resonant growth and can safely be ignored at longer times. Consequently, for very short times, the perturbative calculation more accurately tracks the behavior of the correction. Nonetheless, as the multiple-scales solution demonstrates, the maximum value of the correction increases at times that lie beyond the perturbative regime, highlighting the benefit of analyzing the dynamics over longer time intervals.

\section{Discussion}

\begin{table*}[t]
    \centering
    \begin{tabularx}{\textwidth}{@{\extracolsep{\fill}} lccccc }
        \hline\hline
        \rule{0pt}{10pt}Platform & $\omega_0/2\pi$ & $L_0$ & $\omega/2\pi$ & $\mathcal{A}$ & $N_\mathrm{cyc}$ \\[2pt] \hline
        \rule{0pt}{10pt}Optical cavity QED & $\sim10^{14}$ Hz & $\sim10^{-6}$ m & $\sim10^6$ Hz & $\lesssim10^{-10}$ & $\sim 1$\\
        \rule{0pt}{10pt}Microwave cavity QED & $\sim10^{10}$ Hz & $\sim10^{-2}$ m & $\sim10^6$ Hz & $\lesssim10^{-6}$ & $\sim10$--$10^3$\\
        \rule{0pt}{10pt}Circuit QED & $\sim10^{10}$ Hz & $\sim10^{-2}$ m & $\sim10^8$ Hz & $\lesssim10^{-4}$ & $\sim10^3$ \\[2pt]
        \hline\hline
    \end{tabularx}
    \caption{Relevant parameters in optical cavities, microwave cavities and circuit QED: typical cavity mode frequency $\omega_0$, typical (effective) cavity length $L_0$, highest experimentally achievable modulation frequency $\omega$ compatible with the adiabatic condition $\omega L_0 / c \ll 1$, resulting upper bounds on the modulation amplitude $\mathcal{A}$ required by the perturbative condition $\mathcal{A}\omega_0/\omega\ll1$ (taking $10^{-2}$ as the threshold for $\ll 1$), and highest achievable number of coherent modulation cycles $N_\mathrm{cyc}$ determined by platform-specific limiting factors, as discussed in the main text.}
    \label{tab:platforms}
\end{table*}

The model introduced in this work relies on three key ingredients: (1) Jaynes--Cummings-type light--matter coupling, (2) adiabatic separation, $\omega L_0/c\ll1$, and (3) a tunable cavity frequency with small modulation depth, $\mathcal{A}\ll1$. The modulation effect is strongest near $\Omega_n\approx\omega$, so the highest sensitivity is achieved when $\delta_0$ and $g_n$ are tuned such that $\Omega_n$ lies within the accessible modulation band. Unlike the idealized Jaynes--Cummings model, any experimental atom--cavity platform is an open system with two main loss channels: cavity decay at rate $\kappa$ and free-space atomic polarization decay at rate $\gamma$. Because our model requires coherent atom--field dynamics over a large number $N_\mathrm{cyc}$ of modulation cycles, these dissipative rates must remain smaller than the modulation rate, $\omega\gg\kappa,\gamma$. An overview of relevant experimental platforms is presented in Table~\ref{tab:platforms}.

The most literal implementation of our model is an optical cavity with compliant boundaries (a moving mirror or membrane) and a two-level atom inside. Optical resonances satisfy $\omega_0/2\pi\sim10^{14}$~Hz, with cavity lengths $L_0\sim10^{-7}$--$10^{-5}$~m~\cite{Aspelmeyer2014}. Mechanical modulation frequencies are usually in the kHz--MHz band~\cite{Aspelmeyer2014,Thompson2008,Chan2011}, so the adiabatic condition is safely fulfilled. For this type of modulation, first-order perturbation theory implies an upper bound of $\mathcal{A}\sim10^{-13}$--$10^{-10}$. In this regime, the displacement amplitude $x_0=\mathcal{A}L_0$ becomes extremely small, $x_0\sim10^{-20}$--$10^{-15}$~m. For such displacements, quantum fluctuations are non-negligible, with the fundamental limit set by zero-point motion. A typical scale of quantum fluctuations in nanomechanical systems is $\sim10^{-15}$~m~\cite{Gu2013}. This indicates that a complete description of the optical setup would require a fully quantum treatment of the mirror~\cite{Holz2015}. Additionally, in current optomechanical cavity-QED devices, $\kappa/2\pi$ is commonly in the $10^{5}$--$10^{7}$~Hz range~\cite{Brennecke2007,Colombe2007,Thompson2008,Aspelmeyer2014}, which strongly limits the usable interrogation window and realistically allows only a few coherent modulation cycles, $N_\mathrm{cyc} \sim 1$.

The stringent limitations of optical cavity QED can be relaxed to some extent by considering instead a microwave cavity with resonant frequency $\omega_0/2\pi \sim 10^{10}$~Hz and typical length $L_0 \sim 10^{-2}$~m~\cite{Raimond2001}. Although microwave cavities with quasi-static piezo-tunable lengths have been realized~\cite{Suleymanzade2020}, high-speed periodic boundary modulation has not yet been reported. Thus, implementing our model would require the development of a new experimental setup, which is possible in principle but could reduce the cavity quality factor. Mechanical modulation frequencies still fulfill the adiabatic condition, requiring $\mathcal{A}\sim 10^{-9}$--$10^{-6}$ for the perturbative expansion to remain valid. This gives $x_0 \sim 10^{-11}$--$10^{-8}$~m, well above the zero-point motion and within the range accessible to piezoelectric actuators. 

In standard microwave cavity QED experiments, Rydberg atoms passing through the cavity act as two-level systems. Since the atomic decay rate for circular Rydberg atoms reaches as low as $\gamma/2\pi \sim 10$~Hz~\cite{Richaud2023}, and the lowest achieved cavity decay rates are $\kappa/2\pi \sim 1$~Hz~\cite{Kuhr2007}, the atom--field interaction time is limited predominantly by the atom velocity and the field beam waist, which lead to an upper bound $t \lesssim 100$~\textmu s~\cite{Assemat2019}. Therefore, for modulation in the MHz band, the maximal number of coherent cycles is $N_\mathrm{cyc} \sim 10$. Recent developments include the coherent coupling of optically trapped Rydberg atoms to planar superconducting resonators~\cite{Kaiser2022} and optically accessible high-finesse millimeter-wave cavities designed for trapped Rydberg-atom arrays, with linewidths as small as $\kappa/2\pi \sim 10$~Hz~\cite{Zhang2025}. For modulation in the MHz band, the latter linewidth would permit on the order of $N_\mathrm{cyc} = \omega/(2\pi\kappa) \sim 10^3$ coherent cycles.
 
Superconducting circuit QED is experimentally the most favorable platform for our protocol. There, cavity-frequency modulation can be implemented not only via mechanical motion but also through fast electrical tuning of effective boundary conditions. Microwave resonators with $\omega_0/2\pi\sim10^9$--$10^{10}$~Hz and an effective electrical length $L_0\sim1$~cm are standard~\cite{Wallraff2004,Blais2021}. Using flux-tunable superconducting resonators and rapidly modulated SQUID boundaries, modulation frequencies of up to $\sim10$~GHz are achievable~\cite{Sandberg2008,Wilson2011}; however, the adiabatic condition requires $\omega/2\pi\lesssim100$~MHz. Modulation amplitudes in the conservative range $\mathcal{A}\sim10^{-6}$--$10^{-4}$ remain perturbative while still being experimentally realistic: fast superconducting-resonator tuning experiments report fractional frequency modulation of up to $10^{-1}$~\cite{Sandberg2008,Wilson2011}. Importantly, the cavity linewidths can be engineered from $\kappa/2\pi\sim10^{3}$ to $10^{6}$~Hz~\cite{Blais2021,Paik2011,Reagor2016}, while polarization decay rates for transmon qubits reach as low as $\gamma/2\pi\sim 100$~Hz~\cite{Tuokkola2025}. In this window, the number of coherent modulation cycles is approximately $N_\mathrm{cyc} \sim 1$--$10^3$, indicating that circuit-QED platforms are realistic, state-of-the-art candidates for observing the predicted modulation-induced correction.

For the effect of boundary modulation to be detectable through the proposed mechanism, the measurement uncertainty must satisfy $\delta \mathcal{A} \lesssim \mathcal{A}$, which implies $\mathcal{A} \gtrsim 1/\sqrt{M \mathcal{I}_\mathrm{C}(t)}$. This sets a lower bound on the number $M$ of experimental repetitions required for detection, namely $M \gtrsim (\mathcal{A}^2 \mathcal{I}_\mathrm{C}(t))^{-1}$. For platforms where the cavity linewidth is the largest decay rate, the interrogation time is bounded by $t \lesssim 1/\kappa$, and the maximal achievable classical Fisher information scales as the square of the cavity quality factor, $Q = \omega_0/\kappa$. Consequently, the minimal number of repetitions required for detection scales as $M \sim (Q\mathcal{A})^{-2}$. For superconducting microwave resonators used in circuit QED, the quality factor typically lies in the range $Q \sim 10^4$--$10^7$. Thus, for experimentally realistic modulation amplitudes within the valid perturbative regime ($Q\mathcal{A} \lesssim 1$), only very few measurements are needed, approaching $M \sim 1$ near the validity threshold.

\section{Conclusion}

We have shown that the effect of a plane GW on atom--field dynamics in free space, arising from the modification of the field mode structure, can be simulated through modulated boundary conditions in a strongly coupled atom--cavity system. The derived correction to the atomic transition probability parallels GW imprints on the expected number of photons emitted into a fixed field mode. Our results indicate that the setup required to observe the predicted effect may be within reach of current technology. This opens the possibility of studying simulated GW effects in a controlled setting and identifying opportunities and challenges before devising methods for quantum sensing of genuinely general relativistic phenomena.

\section{Acknowledgments}

J.P., N.A., and M.Z. acknowledge the
Knut and Alice Wallenberg Foundation through a Wallenberg Academy Fellowship No. 2021.0119. P.M. acknowledges the financial support provided by the European Union under the Erasmus+ Program.

\bibliography{bibliography}

\onecolumngrid

\appendix

\section{Perturbative solution}\label{Appendix: Perturbative solution}

In this appendix we derive perturbative solution for the atomic population. As discussed in the main text, we consider a system consisting of a two-level atom coupled to a single radiation mode, which evolves under the Hamiltonian:
\begin{align}
  \hat{H}(t) = \hbar \delta(t) \frac{\hat{\sigma}_z}{2} + \hbar g_0 \left( \hat{\sigma}_+\hat{a} + \hat{\sigma}_-\hat{a}^\dag\right).
\end{align}
Here, $\delta(t) = \delta_0 + \mathcal{A} \omega_0 \cos(\omega t+\phi)$ is the instantaneous atom--field detuning, $\hat{\sigma}_z$, $\hat{\sigma}_+$, and $\hat{\sigma}_-$ are the atomic inversion, raising, and lowering operators, while $\hat{a}$ and $\hat{a}^\dag$ are the bosonic annihilation and creation operators, respectively. The Hamiltonian splits into the time-independent part $\hat{H}_0$ and the time-dependent perturbation Hamiltonian $\hat{V}(t)$, given by:
\begin{equation}
  \hat{V}(t) = \mathcal{A} \hbar \frac{\hat{\sigma}_z}{2} \omega_0 \cos(\omega t+\phi).
\end{equation}

Assume an initial atom--field state of the form $\ket{\psi_0} = \ket{g,n}$, with the atom in the ground state $\ket{g}$ and the field in the $n$-photon state $\ket{n}$. Then, we can restrict the Hamiltonian to the invariant subspace spanned by $\{\ket{e,n-1},\ket{g,n}\}$. Defining the Pauli matrices $\hat{X}$, $\hat{Y}$ and $\hat{Z}$ in this basis, the Hamiltonian $\hat{H}_0$ and the perturbation part $\hat{V}(t)$ take the form:
\begin{align}
    &\hat{H}^{(n)}_0 = \hbar \begin{pmatrix}
        \delta_0/2 & g_n \\
        g_n & -\delta_0/2
    \end{pmatrix} = \frac{\hbar}{2} \left( \delta_0 \hat{Z} + 2g_n \hat{X} \right),\\
    &\hat{V}^{(n)}(t) = \mathcal{A} \hbar \begin{pmatrix}
        \omega_0/2 & 0 \\
        0 & -\omega_0/2
    \end{pmatrix} \cos(\omega t+\phi) = \mathcal{A} \frac{\hbar}{2} \omega_0 \cos(\omega t+\phi) \hat{Z}.
\end{align}
Using the standard Pauli vector exponential identity, we obtain the unperturbed evolution operator:
\begin{equation}
    \hat{U}^{(n)}_0(t) = e^{-\frac{i}{\hbar} \hat{H}^{(n)}_0 t} = \cos(\Omega_n t/2) \hat{\mathbb{I}} - i \sin(\Omega_n t/2) \left(\frac{\delta_0}{\Omega_n} \hat{Z} + \frac{2g_n}{\Omega_n} \hat{X} \right).
\end{equation}
With the free evolution solved, we now perform the perturbative expansion. Up to first order in $\mathcal{A}$, the state of the system at time $t$ is given by:
\begin{equation}
    \ket{\psi(t)} = \hat{U}_0(t) \ket{\psi_0} - \frac{i}{\hbar} \hat{U}_0(t) \int_0^t \der t'\, \hat{V}_\mathrm{I}(t') \ket{\psi_0},
\end{equation}
where $\hat{V}_\mathrm{I}(t)$ is the perturbation Hamiltonian in the interaction picture. The unperturbed transition amplitude $\tilde{c}_e(t)$ is determined by the first term in the expansion:
\begin{equation}
    \tilde{c}_e(t) = \bra{e,n-1} \hat{U}_0(t) \ket{g,n} = - i \frac{2g_n}{\Omega_n} \sin(\Omega_n t/2).
\end{equation}
The first-order correction $\delta c_e(t)$ is given by:
\begin{align}
    \delta c_e(t) &= - \frac{i}{\hbar} \int_0^t \der t'\, \bra{e,n-1} \hat{U}_0(t-t') \hat{V}(t') \hat{U}_0(t') \ket{g,n} \nonumber\\
    &= -\mathcal{A} \omega_0 \frac{2 g_n}{\Omega_n } \int_0^t \der t'\, \left\{\frac{1}{2}\sin[\Omega_n(t'-t/2)] + i \frac{\delta_0}{\Omega_n} \sin(\Omega_n t'/2) \sin[\Omega_n (t'-t)/2] \right\} \cos(\omega t'+\phi). \label{eq:amplitude_correction}
\end{align}
We want to calculate the transition probability $p_e(t) = \abs{c_e(t)}^2$, where $c_e(t) = \tilde{c}_e(t) + \delta c_e(t)$, up to first order in $\mathcal{A}$. Since the unperturbed amplitude $\tilde{c}_e(t)$ is purely imaginary, the probability splits, according to $p_e(t) = \tilde{p}_e(t) + \delta p_e(t)$, into an equilibrium contribution $\tilde{p}_e(t) = \abs{\tilde{c}_e(t)}^2$ and a first-order correction $\delta p_e(t) = 2\Re[\tilde{c}_e^*(t)\delta c_e(t)] = 2 \Im[\tilde{c}_e(t)]\Im[\delta c_e(t)]$. Therefore, it suffices to calculate the imaginary part of $\delta c_e(t)$. To this end, we change
the integration variable in Eq.~\eqref{eq:amplitude_correction} from $t'$ to $u = t'-t/2$ and evaluate the following integral:
\begin{align}
    I &\equiv \int_{-t/2}^{t/2} \der u\, \sin[\Omega_n (u/2+t/4)] \sin[\Omega_n (u/2-t/4)] \left[\cos(\omega u) \cos(\omega t/2+\phi) - \sin(\omega u) \sin(\omega t/2+\phi)\right] \nonumber\\
    &= \int_{-t/2}^{t/2} \der u\, \frac{1}{2}\left[\cos(\Omega_n t/2) - \cos(\Omega_n u) \right] \cos(\omega u) \cos(\omega t/2+\phi) \nonumber\\
    &= \frac{1}{2} \cos(\omega t/2+\phi) \left\{\frac{2 \cos(\Omega_n t/2) \sin(\omega t/2)}{\omega} - \frac{\sin[(\Omega_n - \omega) t/2]}{\Omega_n - \omega} - \frac{\sin[(\Omega_n + \omega) t/2]}{\Omega_n + \omega}\right\} \nonumber\\
    &= \frac{\Omega_n t}{4\omega} \cos(\omega t/2+\phi) \left\{\sinc[(\Omega_n + \omega) t/2] - \sinc[(\Omega_n - \omega) t/2]\right\},
\end{align}
where we use $\sinc(x) \equiv \sin(x)/x$. Substituting back into Eq.~\eqref{eq:amplitude_correction} and introducing the dimensionless parameter $r \equiv \delta_0/\Omega_n$, this gives:
\begin{equation}
    \Im[\delta c_e(t)] = -\frac{1}{2} \mathcal{A} \frac{\omega_0}{\omega} \frac{r \sqrt{1-r^2}}{2} \cos(\omega t/2+\phi)\, \Omega_n t \left\{\sinc[(\Omega_n + \omega) t/2] - \sinc[(\Omega_n - \omega) t/2]\right\}.
\end{equation}
Therefore, the equilibrium contribution $\tilde{p}_e(t)$ and the perturbative correction $\delta p_e(t)$ are given by:
\begin{align}
    &\tilde{p}_e(t) = \left(1-r^2\right) \sin^2(\Omega_n t/2),\\
    &\delta p_e(t) = \mathcal{A} \frac{\omega_0}{\omega} \frac{r(1-r^2)}{4} \cos(\omega t/2+\phi)\, \Omega_n^2 t^2 \sinc(\Omega_n t/2) \left\{\sinc[(\Omega_n + \omega) t/2] - \sinc[(\Omega_n - \omega) t/2]\right\}.
\end{align}
Introducing the functions:
\begin{align}
    &f(\Omega_n,t) \equiv  \Omega_n^2 t^2 \sinc(\Omega_n t/2) \{\sinc[(\Omega_n + \omega)t/2] -\sinc[(\Omega_n - \omega)t/2]\},\\
    &g(r) \equiv r (1-r^2)/4,
\end{align}
we can write the correction in a more compact form:
\begin{equation}
    \delta p_e(t) = \mathcal{A} \frac{\omega_0}{\omega} \cos(\omega t/2+\phi) f(\Omega_n,t) g(r).
\end{equation}
Notice that the correction can be optimized by tuning the parameter $r$ which enters through the function $g(r)$.

\section{Fisher information}\label{Appendix: Fisher information}

Here we present details of the derivation and discussion of the conditions maximizing the classical Fisher information available from the internal atomic state about the mirror’s oscillation amplitude $\mathcal{A}$. As shown in Sec.~\ref{sec:amplitude_estimation}, in the resonant limit $\Omega_n \to \omega$, the classical Fisher information associated with the atomic excited-state population measurement is given by:
\begin{align}\label{eq:Fisher_information}
    \mathcal{I}_\mathrm{C}(t) ={} & \frac{1}{4} \left(\frac{\omega_0}{\omega}\right)^2 h(r,\phi,t)\,[\sin(\omega t) - \omega t]^2,
\end{align}
where $h(r,\phi,t)$ is the modulating function, defined as:
\begin{equation}
    h(r,\phi,t) \equiv \frac{r^2\left(1-r^2\right)\cos^2(\omega t/2 + \phi)}{1-\left(1-r^2\right)\sin^2(\omega t/2)}.
\end{equation}
To extract maximal information about the amplitude $\mathcal{A}$, measurements on the atom should be performed at times corresponding to local maxima of the Fisher information $\mathcal{I}_\mathrm{C}(t)$. For the full expression~\eqref{eq:Fisher_information}, determining optimal interrogation times analytically is impossible; however, approximations can be found by neglecting the monotonic envelope and maximizing the function $h(r,\phi,t)$ over time. To this end, we rewrite the function in the form:
\begin{align}\label{eq:h_rewritten}
    h(r,\phi,t) &= r^2\left(1-r^2\right)\frac{\left[\cos(\phi)\cos(\omega t/2) - \sin(\phi)\sin(\omega t/2)\right]^2}{\cos^2(\omega t/2) + r^2 \sin^2(\omega t/2)} = r^2\left(1-r^2\right)\cos^2(\phi) \frac{\left[1 - \tan(\phi)\tan(\omega t/2)\right]^2}{1 + r^2 \tan^2(\omega t/2)}.
\end{align}
Differentiating this expression with respect to $\xi \equiv \tan(\omega t/2)$, requiring the derivative to vanish and discarding the minimum branch yields the solution $\xi = - \tan(\phi)/r^2$. Thus, the local maxima occur where $\tan(\omega t/2) = - \tan(\phi)/r^2$, which gives:
\begin{equation}
    t_m = 2\left\{\pi m - \arctan[\tan(\phi)/r^2]\right\}/\omega, 
\end{equation}
for $m \in \mathbb{N}$. Substituting these approximate optimal interrogation times back into Eq.~\eqref{eq:h_rewritten} determines the local maxima:
\begin{equation}
    h(r,\phi,t_m) = (1-r^2)[r^2\cos^2(\phi)+\sin^2(\phi)].
\end{equation}
These values can be further optimized by tuning the ratio $r \equiv \delta_0/\Omega_n$ or the initial modulation phase $\phi$, depending on specific experimental capabilities.

It has to be emphasized that we focus on the classical Fisher information rather than the quantum Fisher information because the proposed detection protocol is based on state-selective population readout of the atom. The quantum Fisher information would optimize over all possible measurements on the atom--field state and would therefore provide a measurement-independent upper bound, generally attainable only with phase-sensitive measurements in a basis that depends on the full perturbed evolution. Such an optimization is not directly tied to the experimentally accessible observable considered here. By contrast, $\mathcal{I}_\mathrm{C}(t)$ quantifies the information available from the actual binary measurement and therefore gives the relevant sensitivity estimate for the protocol.

\section{Multiple-scales analysis}\label{Appendix: Multiple-scales analysis}

Here we provide a detailed derivation of the system’s dynamics governed by the Hamiltonian in Eq.~\eqref{eq:Hamiltonian}, employing the multiple-scales approach. Restricting $\hat{H}(t)$ to the $n$-excitation subspace and keeping the notation from Sec.~\ref{sec:perturbative_solution} we obtain:
\begin{align}
  \hat{H}^{(n)}(t) ={}& \hbar \begin{pmatrix}
    \delta(t)/2 & g_n \\
    g_n & -\delta(t)/2
  \end{pmatrix}.
\end{align}
We can decompose any state in the $n$-excitation subspace as:
\begin{equation}
  \ket{\psi(t)} = c_e(t) \ket{e,n-1} + c_g(t) \ket{g,n}.
\end{equation}
The evolution of the system is then governed by the Schr\"odinger equation
\begin{equation}
  i\hbar\partial_t\ket{\psi(t)} = \hat{H}^{(n)}(t)\ket{\psi(t)},
\end{equation}
which results in the following set of differential equations for the amplitudes $c_e(t)$ and $c_g(t)$:
\begin{subequations}\label{eq:cg_ce_system}
\begin{empheq}[left=\empheqlbrace]{align}
  i \dot{c}_e(t) &= \tfrac{1}{2} \delta(t) c_e(t) + g_n c_g(t),\label{eq:cg_ce_system_a}\\
  i \dot{c}_g(t) &= -\tfrac{1}{2} \delta(t) c_g(t) + g_n c_e(t).\label{eq:cg_ce_system_b}
\end{empheq}
\end{subequations}
These equations can be decoupled to yield second-order ordinary differential equations. For concreteness, let us focus on the $c_e$ amplitude, for which we obtain:
\begin{equation}\label{eq:ce_ODE}
  \ddot{c}_e(t) + \frac{1}{4}\left[4 g_n^2 + \delta^2(t) + 2 i \dot{\delta}(t)\right] c_e(t) = 0.
\end{equation}
The time-dependent coefficient in parentheses can be written as:
\begin{align}
  \Omega^2(t) ={} & 4 g_n^2 + \delta_0^2 + 2 \mathcal{A} \omega_0 \delta_0 \cos(\omega t + \phi) - 2 i \mathcal{A} \omega_0 \omega \sin(\omega t + \phi) + \mathcal{A}^2 \omega_0^2 \cos^2(\omega t + \phi).
\end{align}
Recall that $\Omega_n^2 = 4g_n^2 + \delta_0^2$ and introduce a parameter $\alpha \equiv \mathcal{A}\omega_0/\Omega_n$. Since $\Omega_n \approx \omega$, we have $\alpha \ll 1$ consistently with the assumptions from Sec.~\ref{sec:perturbative_solution}. Therefore, up to first order in $\alpha$, we obtain:
\begin{equation}
  \Omega^2(t) = \Omega_n^2 \left[1 + \alpha \frac{\delta_0 - \omega}{\Omega_n} e^{i (\omega t + \phi)} + \alpha \frac{\delta_0 + \omega}{\Omega_n} e^{-i (\omega t + \phi)} \right].
\end{equation}
We now apply the method of multiple-scales analysis. Introduce the slow time scale $t_1 = \alpha t$ and assume the solution $c_e(t)$ is a perturbation-series solution dependent both on $t$ and $t_1$, treated as:
\begin{equation}\label{eq:perturbative_expansion}
  c_e(t) = c_{0}(t,t_1) + \alpha c_{1}(t,t_1) + \mathcal{O}(\alpha^2).
\end{equation}
The time derivative now becomes:
\begin{equation}
  \frac{\der}{\der t} = \frac{\partial}{\partial t} + \frac{\der t_1}{\der t} \frac{\partial}{\partial t_1} = \frac{\partial}{\partial t} + \alpha \frac{\partial}{\partial t_1}.
\end{equation}
Then the zeroth- and first-order problems of the multiple-scales perturbation series for Eq.~\eqref{eq:ce_ODE} become:
\begin{align}
  \frac{\partial^2 c_0}{\partial t^2} + \frac{\Omega_n^2}{4} c_0 &= 0,\\
  \frac{\partial^2 c_1}{\partial t^2} + \frac{\Omega_n^2}{4} c_1 &= - \frac{\Omega^2(t) - \Omega_n^2}{4\alpha} c_0 - 2 \frac{\partial^2 c_0}{\partial t\, \partial t_1}.\label{eq:c1}
\end{align}
The zeroth-order problem has the general solution:
\begin{equation}\label{eq:zeroth-order_solution}
  c_0(t,t_1) = A(t_1) e^{i \Omega_n t/2} + B(t_1) e^{-i \Omega_n t/2}.
\end{equation}
This determines the forcing in the right hand side of the differential equation for the first-order problem. The occurrence of secular terms can be prevented if the terms proportional to $e^{\pm i \Omega_n t/2}$ in the forcing vanish. Introducing $\Omega_n = \omega + \alpha \sigma$, this imposes the following solvability conditions on $A(t_1)$ and $B(t_1)$:
\begin{subequations}\label{eq:A_B_system}
\begin{empheq}[left=\empheqlbrace]{align}
  A'(t_1) &= \tfrac{i}{4}(\delta_0 - \omega) B(t_1) e^{-i(\sigma t_1-\phi)},\label{eq:A_B_system_A}\\
  B'(t_1) &= -\tfrac{i}{4}(\delta_0 + \omega) A(t_1) e^{i(\sigma t_1-\phi)}.\label{eq:A_B_system_B}
\end{empheq}
\end{subequations}
Decoupling these equations, we obtain two second-order ordinary differential equations:
\begin{align}
  A''(t_1) + i \sigma A'(t_1) - \frac{1}{16}\left(\delta_0^2 - \omega^2\right) A(t_1) &= 0,\label{eq:A_B_ODE_A}\\
  B''(t_1) - i \sigma B'(t_1) - \frac{1}{16}\left(\delta_0^2 - \omega^2\right) B(t_1) &= 0.\label{eq:A_B_ODE_B}
\end{align}
The solutions to these equations are:
\begin{align}
  A(t_1) &= e^{-i \sigma t_1/2} \left[A_+ e^{i \nu t_1/2} + A_- e^{-i \nu t_1/2}\right],\label{eq:A_solution}\\ 
  B(t_1) &= e^{i \sigma t_1/2} \left[B_+ e^{i \nu t_1/2} + B_- e^{-i \nu t_1/2}\right]\label{eq:B_solution},
\end{align}
where $\nu = \sqrt{\sigma^2 + \left(\omega^2-\delta_0^2\right)/4}$, and the constants $A_\pm$ and $B_\pm$ are determined by the initial conditions.

At zeroth order in $\alpha$, substituting Eq.~\eqref{eq:zeroth-order_solution} into Eqs.~\eqref{eq:perturbative_expansion} and~\eqref{eq:cg_ce_system_a} at $t=0$ gives:
\begin{align}
  A(0) &= \frac{1}{2} \left[ \left(1 - \frac{\delta_0}{\Omega_n}\right) c_e(0) - \frac{2 g_n}{\Omega_n} c_g(0) \right],\\
  B(0) &= \frac{1}{2} \left[ \left(1 + \frac{\delta_0}{\Omega_n}\right) c_e(0) + \frac{2 g_n}{\Omega_n} c_g(0) \right],
\end{align}
where $A(0)=A_++A_-$ and $B(0)=B_++B_-$. Moreover, using the solvability conditions~\eqref{eq:A_B_system_A}--\eqref{eq:A_B_system_B} at $t_1=0$ one finds
\begin{align}
  A'(0) &= \tfrac{i}{4}(\delta_0-\omega)\,B(0) e^{i\phi},\\
  B'(0) &= -\tfrac{i}{4}(\delta_0+\omega)\,A(0) e^{-i\phi}.
\end{align}
Differentiating Eqs.~\eqref{eq:A_solution}--\eqref{eq:B_solution} and setting $t_1 = 0$ yields
\begin{align}
  A'(0) &= -i\tfrac{\sigma}{2}(A_++A_-)+i\tfrac{\nu}{2}(A_+-A_-),\\
  B'(0) &= i\tfrac{\sigma}{2}(B_++B_-)+i\tfrac{\nu}{2}(B_+-B_-).
\end{align}
Finally, combining the above equalities, we obtain:
\begin{align}
  A_{\pm} &= \frac{1}{2}\left[A(0) \pm \frac{\frac{\delta_0-\omega}{2} B(0) e^{i\phi} + \sigma A(0)}{\nu}\right],\\
  B_{\pm} &= \frac{1}{2}\left[B(0) \mp \frac{\frac{\delta_0+\omega}{2} A(0) e^{-i\phi}+ \sigma B(0)}{\nu}\right].
\end{align}
Substituting for $A(0)$ and $B(0)$, the constants $A_{\pm}$ and $B_{\pm}$ are thus fully expressed in terms of $c_e(0)$ and $c_g(0)$:
\begin{align}
  A_{\pm} ={}& \frac{1}{4}\left[\left(1 - \frac{\delta_0}{\Omega_n}\right) c_e(0) - \frac{2 g_n}{\Omega_n} c_g(0)\right] \nonumber\\
  &\pm \frac{1}{2\nu}\Bigg\{\frac{\delta_0-\omega}{4}\left[\left(1 + \frac{\delta_0}{\Omega_n}\right) c_e(0) + \frac{2 g_n}{\Omega_n} c_g(0)\right] e^{i\phi} + \frac{\sigma}{2}\left[\left(1 - \frac{\delta_0}{\Omega_n}\right) c_e(0) - \frac{2 g_n}{\Omega_n} c_g(0)\right]\Bigg\},\label{eq:A_pm}
  \\
  B_{\pm} ={}& \frac{1}{4}\left[\left(1 + \frac{\delta_0}{\Omega_n}\right) c_e(0) + \frac{2 g_n}{\Omega_n} c_g(0)\right]\nonumber\\
  &\mp \frac{1}{2\nu}\Bigg\{\frac{\delta_0+\omega}{4}\left[\left(1 - \frac{\delta_0}{\Omega_n}\right) c_e(0) - \frac{2 g_n}{\Omega_n} c_g(0)\right] e^{-i\phi} + \frac{\sigma}{2}\left[\left(1 + \frac{\delta_0}{\Omega_n}\right) c_e(0) + \frac{2 g_n}{\Omega_n} c_g(0)\right]\Bigg\}. \label{eq:B_pm}
\end{align}
At leading order in $\alpha$, the excited-state amplitude is
\begin{equation}
  c_e(t) = c_0(t,t_1)\big|_{t_1=\alpha t}
  = e^{i \Omega_n t/2}A(\alpha t) + e^{-i \Omega_n t/2}B(\alpha t).
\end{equation}
In the near-resonant regime considered here, we have $\nu^2 = \sigma^2 + (\omega^2 - \delta_0^2)/4 > 0$, and therefore $\nu\in\mathbb{R}$, so that
\begin{align}
  A(\alpha t) &= e^{-i \sigma \alpha t/2}\left[\left(A_++A_-\right)\cos(\nu\alpha t/2)+i\left(A_+-A_-\right)\sin(\nu\alpha t/2)\right],\\
  B(\alpha t) &= e^{i \sigma \alpha t/2}\left[\left(B_++B_-\right)\cos(\nu\alpha t/2)+i\left(B_+-B_-\right)\sin(\nu\alpha t/2)\right].
\end{align}
Introducing
\begin{align}
  \tilde{A}(\alpha t) &= \left(A_++A_-\right)\cos(\nu\alpha t/2)+i\left(A_+-A_-\right)\sin(\nu\alpha t/2),\label{eq:A_tilde}\\
  \tilde{B}(\alpha t) &= \left(B_++B_-\right)\cos(\nu\alpha t/2)+i\left(B_+-B_-\right)\sin(\nu\alpha t/2),\label{eq:B_tilde}
\end{align}
the leading-order solution can be written as
\begin{equation}
  c_e(t) = \tilde{A}(\alpha t) e^{i \omega t/2} + \tilde{B}(\alpha t) e^{-i \omega t/2},
\end{equation}
where we used $\omega=\Omega_n-\alpha\sigma$. Equivalently,
\begin{equation}\label{eq:c_e_solution}
  c_e(t) = \left[\tilde{A}(\alpha t)+\tilde{B}(\alpha t)\right] \cos\!\left(\omega t/2\right)
  + i \left[\tilde{A}(\alpha t)-\tilde{B}(\alpha t)\right] \sin\!\left(\omega t/2\right).
\end{equation}
For the initial state $\ket{\psi_0} = \ket{g,n}$, i.e. $c_e(0)=0$ and $c_g(0)=1$, Eqs.~\eqref{eq:A_pm}--\eqref{eq:B_pm} simplify to:
\begin{align}
  A_{\pm} ={}& \frac{g_n}{\Omega_n} \left[ - \frac{1}{2} \pm \frac{1}{\nu} \left( \frac{\delta_0-\omega}{4} e^{i\phi} - \frac{\sigma}{2} \right) \right],
  \\
  B_{\pm} ={}& \frac{g_n}{\Omega_n} \left[ \frac{1}{2} \pm \frac{1}{\nu} \left( \frac{\delta_0+\omega}{4} e^{-i\phi} - \frac{\sigma}{2}\right) \right].
\end{align}
Consequently:
\begin{align}
  A_++A_- = -\frac{g_n}{\Omega_n}, \quad &A_+-A_- = \frac{g_n}{\Omega_n} \frac{1}{\nu} \left( \frac{\delta_0-\omega}{2} e^{i\phi} - \sigma \right), \\ 
  B_++B_- = \frac{g_n}{\Omega_n}, \quad &B_+-B_- = \frac{g_n}{\Omega_n} \frac{1}{\nu} \left( \frac{\delta_0+\omega}{2} e^{-i\phi} - \sigma \right).
\end{align}
Finally, substituting these expressions into Eqs.~\eqref{eq:A_tilde}--\eqref{eq:B_tilde} and then into Eq.~\eqref{eq:c_e_solution} gives the explicit leading-order solution:
\begin{align}
  c_e(t) ={}& i \frac{g_n}{\Omega_n} \frac{1}{\nu} \left(\delta_0 \cos(\phi) - i \omega \sin(\phi) - 2 \sigma \right) \sin(\nu \alpha t/2) \cos\!\left(\omega t/2\right) \nonumber\\
  &- i \frac{g_n}{\Omega_n} \left[2 \cos(\nu \alpha t/2) + \frac{1}{\nu} \left(\delta_0 \sin(\phi) + i \omega \cos(\phi) \right) \sin(\nu \alpha t/2) \right] \sin\!\left(\omega t/2\right) \nonumber\\
  ={}& \frac{g_n}{\Omega_n}\frac{1}{\nu} \omega \sin(\nu \alpha t/2) \sin(\omega t/2 + \phi) \nonumber \\
  & + i \frac{g_n}{\Omega_n}\frac{1}{\nu} \left[ \delta_0 \sin(\nu \alpha t/2) \cos(\omega t/2 + \phi) - 2\sigma \sin(\nu \alpha t/2) \cos(\omega t/2) - 2 \nu \cos(\nu \alpha t/2) \sin(\omega t/2)\right].
\end{align}
In the resonant limit $\Omega_n \to \omega$, we have $\nu \to g_n$. Introducing the slow-timescale frequency $\mu \equiv \alpha g_n$ and the ratio $r \equiv \delta_0/\Omega_n$, the amplitude $c_e(t)$ then simplifies to:
\begin{equation}
    c_e(t) = \sin(\mu t/2) \sin(\omega t/2 + \phi) + i r \sin(\mu t/2) \cos(\omega t/2 + \phi) - i \sqrt{1-r^2} \cos(\mu t/2) \sin(\omega t/2).
\end{equation}
Subtracting the equilibrium contribution $\tilde p_e(t) = (1-r^2) \sin^2(\omega t/2)$ from the full probability $p_e(t) = \abs{c_e(t)}^2$ gives the probability correction:
\begin{align}
    \delta p_e(t) ={}& r\sqrt{1-r^2} \sin(\mu t/2) \cos(\mu t/2) \sin(\phi) + r^2 \sin^2(\mu t/2)\nonumber\\
    &+ \left[(1-r^2) \sin^2(\mu t/2) \sin(\phi) - r\sqrt{1-r^2} \sin(\mu t/2) \cos(\mu t/2)\right] \sin(\omega t + \phi).
\end{align}
This expression clearly separates the long- and short-timescale behavior. Defining the long-timescale baseline $\mathcal{B}(r,\phi,t)$ and the envelope $\mathcal{E}(r,\phi,t)$ by
\begin{align}
    &\mathcal{B}(r,\phi,t) \equiv r\sqrt{1-r^2} \sin(\mu t/2) \cos(\mu t/2) \sin(\phi) +r^2 \sin^2(\mu t/2),\\
    &\mathcal{E}(r,\phi,t) \equiv (1-r^2) \sin^2(\mu t/2) \sin(\phi) - r\sqrt{1-r^2} \sin(\mu t/2) \cos(\mu t/2),
\end{align}
the correction can be written compactly as
\begin{equation}
    \delta p_e(t) = \mathcal{B}(r,\phi,t) + \mathcal{E}(r,\phi,t) \sin(\omega t + \phi).
\end{equation}

\end{document}